\documentstyle[prd,aps,psfig]{revtex}
\bibstyle{unsrt}

\tighten
\begin{document}
\draft

\preprint{UNIGE-TH-98-12-1023}
\twocolumn[\hsize\textwidth\columnwidth\hsize\csname 
@twocolumnfalse\endcsname

\title{Predicting the critical density of topological defects 
in $O(N)$ scalar field theories.}
\author{Nuno D. Antunes$^1$, Lu\'{\i}s M. A.  Bettencourt$^2$ and 
Andrew Yates$^3$}
\address{$^1$D\'ept. de Physique Th\'eorique, Universit\'e de Gen\`eve,
24 quai E. Ansermet, CH 1211, Gen\`eve 4, Switzerland}
\address{$^2$T-6/T-11, Theoretical Division, Los Alamos National Laboratory,
Los Alamos NM 87545,USA}
\address{$^3$Centre for Nonlinear Dynamics and its Applications,
University College, Gower Street, London WC1E 6BT, UK}
\date{\today}
\maketitle

\begin{abstract}
$O(N)$ symmetric  $\lambda \phi^4$ field theories 
describe many critical phenomena in the laboratory and in the 
early Universe. Given $N$ and  $D\leq 3$, the dimension of space,
these models exhibit topological defect classical solutions 
that in some cases fully determine their critical behavior. 
For $N=2,~D=3$ it has been observed that   
the defect density is seemingly a universal quantity at $T_c$.
We prove this conjecture and show how to predict its value 
based on the universal critical exponents of the field theory.
Analogously, for general $N$ and $D$ we  
predict the universal critical densities of domain 
walls and monopoles, for which no detailed thermodynamic study exists. 
This procedure can also be inverted, producing an algorithm 
for generating typical defect networks at criticality, 
in contrast to the canonical procedure \cite{VV}, 
which applies only in the unphysical 
limit of infinite temperature. 
\end{abstract}

\pacs{PACS Numbers : 05.70.Fh, 11.27.+d, 98.80.Cq \hfill 
UNIGE-TH-98-12-1023}

\vskip2pc]

$O(N)$ symmetric scalar field theories are a class of models describing 
the critical behavior of an great variety of important physical systems.
For example, for $N=3$ they describe ferromagnets, the
liquid vapor transition and binary mixtures for $N=1$ and 
superfluid $^4He$ and the statistical properties of polymers, for $N=2$.
In the early Universe $N=2$ describes the phase transition associated
with the breakdown of Peccei-Quinn symmetry, and models of high energy
particle physics may belong to the universality class 
of $O(N)$ scalar models, whenever the mass of the Higgs bosons is larger
that that of the gauge bosons. $O(N)$ scalar models are also  
invoked in most implementations of cosmological inflation.  

One of the fundamental properties of $O(N)$ $\lambda \vert \phi \vert^4$ 
field theories is the existence, 
for $N\leq D$, of static non-linear classical solutions, 
(domain walls, vortices, monopoles) that we will refer to 
henceforth as topological defects.
At sufficiently high temperatures, topological defects can be 
excited as non-perturbative fluctuations. Their dominance over the 
thermodynamics, due to their large configurational entropy,  
is known to trigger the phase transition in $O(2)$ in 3D
and 2D, and their persistence at low energies  
prevents the onset of long range order in $O(2)$ $D \leq 2$ 
and in $O(1)$ in 1D.

It is therefore natural that the universal critical exponents
characterizing the phase transition in terms of defects and through 
the behavior of field correlators must be connected.   
This connection is made more quantitative whenever one can construct
dual models, field theories which possess these collective solutions
as their fundamental excitations \cite{Kleinert}. In the absence of 
supersymmetry rigorous mappings between the 
fundamental models and their dual counterparts  
exist only in very special cases \cite{Kleinert,Baxter}. 
Duality has been suggested and empirically observed to be 
a much more general phenomenon, though. 

In this letter we explore the duality between the critical behavior
of the 2-point field correlation function and defect densities 
at criticality.
We will show that it leads to the  proof that the 
critical density of vortex strings, 
observed in recent non-perturbative thermodynamic studies of 
$O(2)$, is a universal number. 
Among other insights \cite{Williams} this shows that 
the phase transition in $O(2)$ in 3D
occurs when a critical density of defects is reached, connecting directly 
the familiar picture of the Hagedorn transition in vortex densities
to the more abstract critical behavior of the fields.
We also extend our procedure to different 
$N$ and $D$, making predictions for the values of the 
universal densities of domain walls and monopoles, in 2 and 3D.

Finally the inversion of this procedure allows us to easily generate 
typical field configurations at criticality. This is of fundamental practical
importance. Recent experiments in $^3He$ \cite{He3} 
and large scale numerical studies of the theory \cite{ABZ} 
have lent quantitative support to the ideas, due to Kibble \cite{K76} and 
Zurek \cite{Z85}, 
that defects form at a second order phase transition due to critical 
slowing down of the fields response over large length scales, in the vicinity
of the critical point. Defect networks hence formed have densities
and length distributions set by thermal equilibrium at $T=T_c^{+}$. 

In contrast most realizations of defect networks used, eg., in 
cosmological studies are generated using the 
Vachaspati-Vilenkin \cite{VV} (VV) algorithm.
It relies on laying down random field phases on a 
lattice and searching for their integer windings 
along closed paths. The absolute 
randomness of the phases corresponds to the $T\rightarrow \infty$ 
limit of the theory. 
More fundamentally it yields defect networks that are 
quantitatively distinct from those in equilibrium at criticality, i.e., 
at formation.   

\begin{figure}
\centerline{\psfig{file=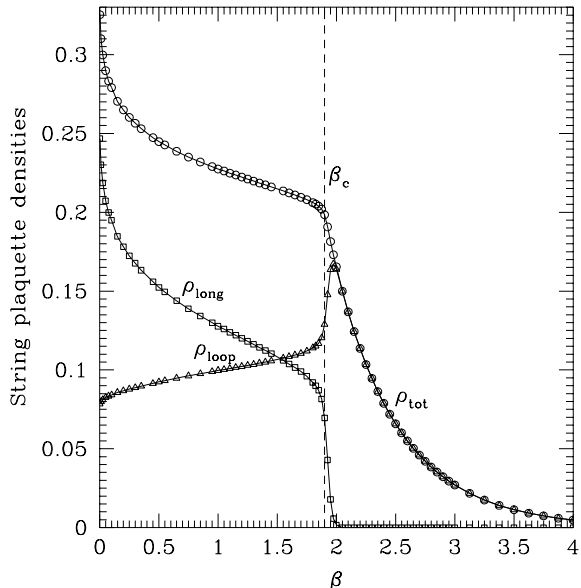,width=3.25in}}
\caption{The string densities, total, loops and long string, as a function
of inverse temperature $\beta$. At $\beta_c$, 
the densities display derivative discontinuities,
signaling a second order phase transition. $\rho_{\rm tot}(\beta_c)$ 
coincides for different studies, leading to the 
conjecture that it is a universal number.}
\label{fig1}
\end{figure}

Fig.~\ref{fig1}  shows the behavior of a
system of vortex strings at a second-order phase-transition, for $O(2)$ 
in 3D. The data was obtained from the study of the non-perturbative 
thermodynamics of the field theory \cite{Us1}.
At $T_c$ the total density of string $\rho_{\rm tot}$
displays a discontinuity in its derivative, signaling a second order 
phase transition.   

A disorder parameter can be constructed in terms of string quantities
by dividing the string population into long string 
(typically string longer than $\sim L^2$, where $L$ is the size of
the computational domain) and loops, comprising of shorter strings.
The corresponding densities are denoted by $\rho_{\rm long}$ and
$\rho_{\rm loop}$. In Fig.~\ref{fig1} we can observe that $\rho_{\rm long}$
consistently vanishes below $T_c$, except for a small
range of $\beta$ where it increases rapidly 
to a finite critical value. In \cite{Us1} we conjectured that
in the infinite volume limit $\rho_{\rm inf}$ exhibits a discontinuous
transition.

The value of the total string density at $\beta_c$,
$\rho_{\rm tot}(\beta_c) \simeq 0.20$ coincides with 
results from studies of different models in the same
universality class \cite{Us1,XY}. This fact lead us to the 
conjecture \cite{Us1} that $\rho_{\rm tot}(\beta_c)$ is
universal.

In order to prove this conjecture we appeal to a well known result, due to
Halperin and Mazenko \cite{Halperin}. 
Halperin's formula expresses $\rho_0$, the density of zeros of a  
Gaussian field distribution in terms of its two-point function. 
For an $O(N)$ theory the relevant quantity is the 
$O(N)$ symmetric correlation function 
$G(x)= \langle \phi(0) \phi(x)^\dagger\rangle$, resulting in
 \begin{equation}
 \rho_0 \propto \left|{G''(x=0)\over G(x=0)}\right|^{N\over 2}. 
 \label{halp} 
 \end{equation}
Eq.~(\ref{halp}) measures the density of coincident zeros 
of all $N$ components of the field at a point. 
Coincident zeros occur at the core of topological defects.
Depending on $N$
and $D$, coincident zeros can be
interpreted as either monopoles, strings or domain walls. 
In the particular case of a Gaussian $O(2)$ theory in $D=3$, 
Halperin's formula allows us to compute the density of vortex
strings crossing an arbitrary plane in three dimensional space, a
quantity that is clearly proportional to 
$\rho_{\rm tot}$.

The last key observation is that in the critical 
domain of a second order transition, all $O(N)$ theories are effectively 
approximately Gaussian,  but with non-trivial
critical exponents. In particular renormalization group analysis shows that 
the mass and quartic coupling vanish at $T_c$ \cite{ZJustin,Wetterich}. 
Higher order polynomial terms (eg. $\propto \phi^6$) may be generated 
but are small. Hence in the critical domain the field two-point 
function can be written as
 \begin{equation}
 G({\bf x}) \propto \int d^{D}{\bf k} \ \frac{e^{i{\bf k}\cdot{\bf x}}}
{|{\bf k}|^{-2 + \eta}}, 
 \label{two-point}
 \end{equation}
  where $\eta << 1.$ is the universal  
critical exponent taking into account deviations
from the mean-field result.
  
Thus the effective Gaussianity of the theory
allows us to use Halperin's result to compute  the critical value 
of $\rho_{\rm tot}(\beta_c)$. Note that, modulo renormalization, 
the final result depends {\it only} on $\eta$
establishing, as conjectured, that $\rho_{\rm tot}(\beta_c)$ is
a universal quantity.
 
 Substituting, Eq.~(\ref{two-point}) 
into (\ref{halp})  we obtain: 
 \begin{equation}
 \rho_{tot} \propto 
 \left({\eta+1\over{\eta+3}}\right)
 {{k_{\rm max}^{3+\eta} - k_{\rm min}^{3+\eta}}
\over k_{\rm max}^{1+\eta} - k_{\rm min}^{1+\eta}},
 \end{equation}
 where we have introduced upper and lower momentum cut-offs 
$k_{\rm max}$ and $k_{\rm min}$. In the case of a lattice
of size $L$ and lattice spacing $a$ we take $k_{\rm min}=2\pi/L$
and $k_{\rm max}=2\pi/a$ which leads to
 \begin{equation}
 \rho_{tot} \propto 
 \left({\eta+1\over{\eta+3}}\right)
 {{1 - (a/L)^{3+\eta}}
 \over 1 - (a/L)^{1+\eta}}.
 \label{halp_scaling}
 \end{equation}
 For large enough lattices, $a/L<<1$, and we obtain
 \begin{equation}
  \rho_{tot} \propto \left({\eta+1\over{\eta+3}}\right).
 \label{scal_theor} 
 \end{equation}
In order to generate quantitative predictions we need to 
determine the exact proportionality factor in Eq.(\ref{scal_theor}). 
 This can be achieved by invoking the other instance when the interacting 
theory becomes Gaussian. In the high temperature limit 
$\beta\rightarrow 0$, the effective interaction becomes irrelevant. 
In terms of renormalization group the theory displays a (trivial) 
Gaussian fixed point, with vanishing effective coupling.
On the lattice fields at different
points will be completely uncorrelated. Note that this situation
corresponds to the usual VV algorithm where
a field is thrown randomly on the lattice and a network of strings
is built by identifying phase windings. Fig.~\ref{fig1}
shows the agreement of the densities observed at $\beta=0.$
with the well known VV result of $\rho_{\rm tot}=1/3$ with 75\% 
long string. Since the totally uncorrelated field corresponds to
a flat power spectrum $G(k)\sim k^0$ we normalize Halperin's
expression by imposing $\rho_{\rm tot} = 1/3$
for $\eta=2$,  
\begin{equation}
\rho_{tot} = {5\over 9} 
\left({\eta+1\over{\eta+3}}\right).
\end{equation}
$\eta$ is always much smaller than $1$. Setting $\eta=0$
we obtain $\rho_{tot}(T_{c})=5/27=0.185$, close to the value 
$\rho_{tot}\simeq 0.2$ observed
both in the $\lambda \phi^{4}$ \cite{Us1} 
and $XY$ \cite{XY} studies in 3D. A more accurate
estimate can be obtained by replacing $\eta$ by its precise value for
the universality class to which these theories belong. From \cite{ZJustin}
we have $\eta\simeq 0.035$ we get $\rho_{\rm tot}=0.190$, 
closer to the observed value. 

 A similar exercise permits the computation of the critical density
of domain walls for $O(1)$ and monopoles in a $O(3)$ theory
at the critical temperature in 3D. 

The  density of domain walls per link is
\begin{equation}
\rho_{tot} = {1\over 2}\left({5 \over 3}\right)^{1/2} 
\left({\eta+1\over{\eta+3}}\right)^{1/2}.
\end{equation}
Note the value of $\rho_{tot}(\beta=0)=1/2$ corresponding to the
high-temperature limit.  
At $\beta_c$ we get, with $\eta=0.034$ \cite{ZJustin},
$\rho_{tot}(\beta_c)\simeq 0.38$.
For monopoles we will take for the flat-spectrum case 
$\rho_{tot}(\beta=0)\simeq 0.1$.
A better estimate can be obtained from a tetrahedral
discretization
of the sphere, resulting in $\rho_{tot}(\beta=0)= 3 / 32$ and 
\begin{equation}
\rho_{tot} = {3\over 32}\left({5 \over 3}\right)^{3/2} 
\left({\eta+1\over{\eta+3}}\right)^{3/2}.
\end{equation}
with $\eta=0.038$ \cite{ZJustin} leading to the critical
value $\rho_{tot}(\beta_c)=0.040$.
Finally for domain walls in 2D, the density per link at $\beta_c$ is
(using $\rho_{tot}(\beta=0)=1/2$):
\begin{equation}
\rho_{tot} = {1\over \sqrt{2}} 
\left({\eta\over{\eta+2}}\right)^{1/2}.
\end{equation}
Taking $\eta=0.26$ \cite{ZJustin} we obtain $\rho_{tot}(\beta_c)=0.24$.

The present procedure can be inverted to generate a typical 
defect network at criticality. 
The approximate 
Gaussianity of the field theory at $T_c$ implies that the statistical 
distribution of fields, $P[\phi]$ is given by
\begin{eqnarray}
P[\phi] = {\cal N} e^{- \beta \int d^3 k {1 \over 2} \phi_{-k} 
G(\vert k \vert) \phi_{k}}
\end{eqnarray}
This distribution can be sampled by generating fields as 
\begin{eqnarray}
\phi_k = R(k) \sqrt{\beta^{-1} G(\vert k \vert )}
\end{eqnarray}
where $R(k)$ is a random number extracted from a Gaussian distribution,
with zero mean and unit variance.
The field can then be Fourier transformed to coordinate space, its phases
identified at each site,   and vortices found in the standard way. 
 Since we will be willing to compare results from this algorithm 
with the ones measured in lattice Langevin simulations we chose to
 employ the exact form of the field correlator on the lattice:
\begin{eqnarray}
G(\vert k \vert)^{-1} = 
\left[ \sum_{i=1}^D 2 \left( 1 - \cos(k_i) \right) \right]^{2-\eta \over 2}
 \sim_{\vert k \vert \rightarrow 0} 
 \vert {\bf k} \vert^{2-\eta}.
 \label{latt-corr}
\end{eqnarray}   
  
\begin{figure}
\centerline{\psfig{file=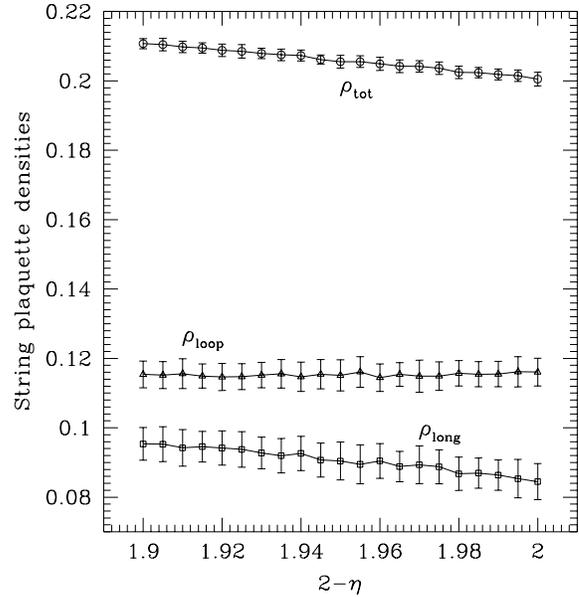,width=3.25in}}
\caption{The string densities from the 50 Gaussian realizations
as a function of $\eta$ for a lattice with $N_{\rm lat}=64$. Error bars 
indicate standard deviation from the mean.}
\label{fig2}
\end{figure}

 We have performed several tests on the algorithm, by comparing it to the 
results of the non-perturbative thermodynamics of the fields at criticality.
We used lattices of size $N_{\rm lat}^3$ with 
$N_{\rm lat}=16,32,64$ and $128$. All 
results are averages over 50 samples obtained from independent random
realizations.  
Fig.~\ref{fig2} shows the string densities for values of $\eta$ between
0. and 0.1, including all reasonable values of $\eta$ in 3D.

\begin{figure}
\centerline{\psfig{file=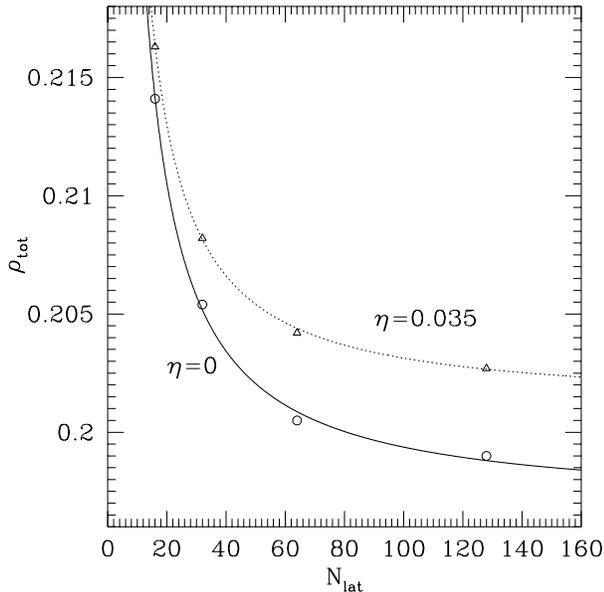,width=3.25in}}
\caption{The total string density for two values of $\eta$ 
 for $N_{\rm lat}=16,32,64$ and $128$ and 
respective fits to a power law. Statistical errors are much larger
than the deviation of the points to the fits.}
\label{fig3}
\end{figure}

 The values for the densities depend on the size of the 
lattice, converging to finite values for large $N_{\rm lat}$.
 In Fig. \ref{fig3} we can see the scaling of $\rho_{\rm tot}$ with
box size for two choices of the critical exponent, the mean field value
$\eta=0.$ and the theoretical result for the $O(2)$ universality class
in 3D, $\eta\sim0.035$. 
We can predict the form and the power of this scaling through 
Eq.~(\ref{halp_scaling}). Writing $a/L=1/N_{\rm lat}$, 
the number of points in the lattice, 
and expanding Eq.~(\ref{halp_scaling}) in powers of $1/N_{\rm lat}$
we see that Halperin's result converges to its infinite volume
limit according to:
 \begin{equation}
 \rho_{\rm tot}(\infty)-\rho_{\rm tot}(N_{\rm lat}) = 
\frac{1}{N_{\rm lat}^{1+\eta}} + O(1/N_{\rm lat}^2)
 \label{eq_scaling}
 \end{equation}
To check these scalings we fitted the data of Fig.~\ref{fig3} 
to a power law of the form:  
 \begin{equation}
 \rho_{\rm tot}(N_{\rm lat}) = \rho_{\rm tot}({\infty}) 
+ \frac{A}{N_{\rm lat}^\alpha}
 \end{equation}
 For $\eta=0$  and $\eta=0.035$ we found:
 \begin{eqnarray}
 &&\eta=0.0,\;\;\;\; 
 \rho_{\rm tot}({\infty})=0.1969,\;\;
 A=0.3259,\;\;
 \alpha=1.060;
 \nonumber \\
 &&\eta=0.035,\;\;\;\; 
 \rho_{\rm tot}({\infty})=0.2012,\;\;
 A=0.3422,\;\;
 \alpha=1.124. \nonumber
 \end{eqnarray}
 These values of $\alpha$ are indeed close to 1,
 with a larger correction for $\eta=0.035$ as expected from
 Eq.~(\ref{eq_scaling}).

  In \cite{Us1}
 for a lattice of size $N_{\rm lat}=100$ we measured $\rho_{\rm tot}(\beta_c)=
 0.198\pm 0.004$. For a Gaussian field with $\eta=0.035$ we obtain
 $\rho_{\rm tot}=0.203\pm 0.003$. The agreement of the two results 
 is very satisfactory.

 The results for $\rho_{\rm long}$ and $\rho_{\rm loop}$ using these
 two different methods are also in good agreement. In this case we 
 were not able to find a reasonable scaling expression though, but 
this is to be expected given the arbitrariness of the long string definition. 
 The results for $N_{\rm lat}=100$, using $\eta=0.035$ are
 $\rho_{\rm long} = 0.080 \pm 0.004$, $\rho_{\rm loop} = 0.121
 \pm 0.004$.
 These compare well with the non-perturbative results
 $\rho_{\rm long} =0.076  \pm 0.005$, $\rho_{\rm loop} = 0.120
 \pm 0.004$.
 Even more impressive is that the string length distribution at criticality
 can also be reproduced by our Gaussian field algorithm. This distribution
can be successfully fitted to an expression of the form \cite{Us1},
\begin{eqnarray}
n(l) = A l^{-\gamma} e^{-\beta \sigma l}. 
\end{eqnarray}
 The fit to the results of the Gaussian field algorithm shows a small 
 variation of the parameters $A, \gamma$, and $\sigma$ for 
 $\eta\sim 0. - 0.1$.
 $\sigma$ is consistently zero, reflecting the fact that the spectrum is 
 always scale invariant. 
 The value of $\gamma$ varies between 2.34 and 2.40.
 For the critical exponent $\eta=0.035$ we obtained $\gamma\simeq 2.35$.
 Once again this is in good agreement with the result from the lattice
 non-perturbative thermodynamics at $T_c$ \cite{Us1}, $\gamma \simeq 2.36$.
 
 Finally the predictions for $\rho_{\rm tot}$ from
 Halperin's formula, when compared to the accuracy of the Gaussian 
 algorithm seem rather poor. The expression is meant to apply for 
 continuum distributions, while all other values of 
 $\rho_{\rm tot}$ were obtained on the lattice. 
 A straight substitution of the lattice correlators (\ref{latt-corr}) into 
 (\ref{halp})
 increases $\rho_{\rm tot}$ to 0.21 from 0.19, covering our full range
 of results. To perform a precise comparison however Halperin's formula
 should be rederived for a field theory on the lattice.
 Despite these shortcomings Halperin's formula
 has the merit of being the only analytic way of estimating the
 critical densities of defects in theories where non-perturbative 
 thermodynamic results are scarce.

 We have therefore established the connection between the universal critical
exponent characterizing the behavior of the $O(N)$ field 2-point correlator 
and the critical density of defects. This
relation implies that defect densities at $T_c$ for a system undergoing a
second order phase transition are universal numbers. 
We predicted them for several $O(N)$ models in 2 and 3D.
Based on these insights we proposed a new algorithm 
for generating networks of defects at the time of formation.   
In particular, we have shown that this algorithm reproduces accurately all the
features of a string network in 3D at criticality. This procedure, instead
of the more usual algorithm of \cite{VV} should be used 
to generate typical defect networks at the time of their formation.

We thank R.~D\"urrer and W.~Zurek for useful discussions.
This work was partially supported by the ESF
and by the DOE, under contract W-7405-ENG-36.

\end{document}